\documentclass[12pt,letterpaper]{article}

\usepackage{amsmath}
\usepackage[psamsfonts]{amssymb}
\usepackage{amsthm}
\usepackage{fullpage}
\usepackage{cite}
\usepackage{indentfirst}
\usepackage{graphicx}
\usepackage{xspace}

\newcommand{\R}{{\mathbb R}}

\newcommand{\ti}[1]{\textit{#1}}
\newcommand{\half}{\ensuremath{\frac{1}{2}}}
\newcommand{\abs}[1]{\lvert#1\rvert}

\newcommand{\dddx}{\frac{d^2}{dx^2}}

\newcommand{\dddrho}{\frac{d^2}{d\rho^2}}
\newcommand{\rn}{Reissner-Nordstr\o{}m\xspace}
\newcommand{\as}{\textrm{\ as\ }}
\newcommand{\im}{\textrm{Im\ }}
\newcommand{\re}{\textrm{Re\ }}
\newcommand{\spt}{\ensuremath{\bigstar}\xspace}
\newcommand{\bH}{\beta}
\newcommand{\bHI}{\beta_I}
\newcommand{\bL}{\beta_L}
\newcommand{\bR}{\beta_R}
\newcommand{\THa}{\ensuremath{T_H}}
\newcommand{\oh}{\hat{\omega}}

\newcommand{\OO}{{\mathcal O}}
\newcommand{\err}{\ensuremath{{\mathcal O}\left(1/\sqrt{\abs{\omega} r_H}\right)}\xspace}
\def\n#1{({\bf{#1}})}

\newcommand{\fig}[2]{
\begin{figure}[t]
\resizebox{6in}{!}{\includegraphics{figures/#1.eps}}
\caption{\label{#1}#2}
\end{figure}
}

\newcommand{\sfigc}[2]{
\begin{figure}
\centering
\resizebox{3in}{!}{\includegraphics{figures/#1.eps}}
\caption{\label{#1}#2}
\end{figure}
}

\makeatletter
\@addtoreset{equation}{section}
\renewcommand\theequation{\thesection.\@arabic\c@equation}
\makeatother

\begin{document}

\bibliographystyle{utphys}

\setcounter{page}{1}
\pagestyle{plain}

\begin{titlepage}

\begin{center}
\hfill HUTP-03/A027\\
\hfill hep-th/0304080

\vskip 1.5 cm
{\huge Greybody factors at large imaginary frequencies}
\vskip 1.3 cm
{\large Andrew Neitzke}\\
\vskip 0.5 cm
{Jefferson Physical Laboratory,
Harvard University,\\
Cambridge, MA 02138, USA}
\vskip 0.3cm
{neitzke@fas.harvard.edu}
\end{center}

\vskip 0.5 cm
\begin{abstract}
Extending a computation which appeared recently in \cite{Motl:2003cd}, we compute the transmission and 
reflection coefficients for massless uncharged scalars and gravitational waves scattered by $d \ge 4$ 
Schwarzschild or $d=4$ \rn black holes, in the limit of large imaginary frequencies.  The transmission
coefficient has an interpretation as the ``greybody factor'' which determines the spectrum
of Hawking radiation.  The result has an interesting structure and we 
speculate that it may admit a simple dual
description; curiously, for \rn the result suggests that this dual description should
involve both the inner and outer horizons.  We also discuss some numerical evidence in favor 
of the formulas of \cite{Motl:2003cd}.
\end{abstract}

\end{titlepage}

\section{Introduction}

Recently there has been a resurgence of interest in certain classical aspects of wave propagation
in black hole backgrounds.  Specifically, it has been suggested that
scattering in the black hole background at large imaginary frequencies may carry information about
quantization of the horizon area \cite{Hod:1998vk,Dreyer:2002vy,Corichi:2002ty,Hod:2003jn}, 
which has led to some recent investigations of the large-imaginary-frequency behavior. 
These recent studies \cite{Motl:2003cd, Motl:2002hd,Berti:2003zu}
mostly focused on the large-imaginary-frequency ``quasinormal modes,'' which are poles of the transmission coefficient 
and reflect the black hole's ringdown response to a perturbation, although \cite{MaassenvandenBrink:2003as}
also recently studied different aspects of the large-imaginary-frequency behavior using methods similar to
those employed in \cite{Motl:2003cd} and in this paper.

In Section \ref{section-transmission-amplitude} of this paper we present a 
computation of the transmission and reflection coefficients for waves traveling
from spatial infinity to the black hole horizon, in the limit of large imaginary frequencies: 
$\omega \to +i \infty$ or more precisely $\abs{\omega r_H} \gg 1$, where $r_H$ is the horizon radius.
We consider here the case of a massless scalar field; then for Schwarzschild 
black holes in $d \ge 4$, the result is\footnote{There is a branch cut in $T(\omega)$ and $R(\omega)$ at positive imaginary frequencies;
to switch to the other side of the cut one replaces $\omega$ by $-\omega$ in \eqref{transmission-schwarzschild},
\eqref{reflection-schwarzschild}, \eqref{transmission-rn-4d}, \eqref{reflection-rn-4d}.}
\begin{align} 
T(\omega) & \approx \frac{e^{\beta \omega} - 1}{e^{\beta \omega} + 3}, \label{transmission-schwarzschild} \\
R(\omega) & \approx \frac{2i}{e^{\beta \omega} + 3}, \label{reflection-schwarzschild}
\end{align}
and for \rn in $d=4$ we get
\begin{align}
T(\omega) & \approx \frac{e^{\beta \omega} - 1}{e^{\beta \omega} + 2 + 3e^{-\bHI \omega}}, \label{transmission-rn-4d} \\
R(\omega) & \approx i\sqrt{3} \frac{1 + e^{-\bHI \omega}}{e^{\beta \omega} + 2 + 3e^{-\bHI \omega}}. \label{reflection-rn-4d}
\end{align}
Here $\bH$ is the inverse Hawking temperature, 
and for \rn $\bHI$ is the inverse Hawking temperature of the
inner horizon; the symbol $\approx$ means that there are \err corrections both to the numerator and to the 
denominator.\footnote{In the case of \rn the coefficient of these 
corrections becomes large in the Schwarzschild
limit, which explains why the results do not simply reduce to those for Schwarzschild
as $\bHI \to 0$.}
The computation is a slight extension of one which recently appeared in \cite{Motl:2003cd}; the
new result is that we obtain the numerator as well as the denominator.  We also calculate $T(\omega)$ and
$R(\omega)$ in the limit $\omega \to -i \infty$.  The results are summarized in Section \ref{section-other-black-holes}.

As we discuss in Section \ref{section-hawking}, the full result has a suggestive structure;
it is strongly reminiscent of the results of \cite{Maldacena:1997ix, Maldacena:1997ih}, which showed that it is
possible to reconstruct the effective SCFT description of near extremal black
holes by studying $T(\omega)$ for those black holes at small real frequencies.  
Essentially the idea is that the scattering
of Hawking radiation back down the hole by the spacetime curvature modifies the blackbody spectrum in a precise
way and makes it agree with the spectrum obtained by an SCFT calculation.
We are led to conjecture the existence of an effective description which 
determines the Hawking radiation at large
imaginary frequencies and involves excitations with rather exotic statistics dictated by
the denominators of \eqref{transmission-schwarzschild}, \eqref{transmission-rn-4d}
(such statistics were also 
proposed in \cite{Motl:2002hd} for $d=4$ Schwarzschild).  We discuss this conjecture in Section \ref{section-hawking}.

Then switching gears, in Section \ref{section-numerical-results} we discuss the 
numerical evidence supporting the analytical computations of \cite{Motl:2003cd} for the denominators, 
some of which emerged only very recently \cite{Berti:2003zu}.
The comparison to the numerical results also clarifies some subtleties about the relation between the
formulas for \rn and for Schwarzschild.

We conclude with a few remarks on prospects for future progress.
In two appendices, we discuss the analytic structure of the transmission and reflection
coefficients and give a few more details of the asymptotic matching procedure used in 
the computation.

\section{The transmission and reflection coefficients} \label{section-transmission-amplitude}

In this section we will compute the transmission and reflection coefficients 
for waves propagating from infinity to the
horizon of a static black hole.  The analysis has been carried out so far for Schwarzschild in $d \ge 4$
and for \rn in $d=4$.  To be concrete we mostly discuss Schwarzschild in $d=4$; in Section 
\ref{section-other-black-holes} we give the results for other cases.

\subsection{The wave equation}

We consider wave propagation in the exterior region of a Schwarzschild black hole in $d=4$,
\begin{equation} \label{metric}
ds^2 = - f(r) dt^2 + f(r)^{-1} dr^2 + r^2 d\Omega^2.
\end{equation}
The coordinates \eqref{metric} run from $r_H < r < \infty$, where
\begin{equation}
r_H = 2GM
\end{equation}
and
\begin{equation}
f(r) = 1-r_H / r.
\end{equation}

The propagating field can be a minimally coupled massless scalar, an electromagnetic test-field, or a linearized perturbation of the metric; we label these three possibilities respectively
by their spins $j=0,1,2$.  In each case the appropriate wave equation is separable.
For example, 
in the $j=0$ case, writing
\begin{equation} \label{solution-separation}
\phi(r,t,\Omega) = r \psi(r) Y_{lm}(\Omega) e^{i \omega t}
\end{equation}
the massless Klein-Gordon equation $\Delta \phi = 0$ becomes
a one-dimensional wave equation
on the interval $-\infty < x < \infty$, deformed by a potential:
\begin{equation} \label{wave-equation}
\left[ - \dddx + V[r(x)] - \omega^2 \right] \psi(x) = 0,
\end{equation}
where
\begin{equation} \label{scalar-potential}
V(r) = f(r) \left[\frac{l(l+1)}{r^2} + \frac{r_H}{r^3} \right]
\end{equation}
and we have introduced the ``tortoise coordinate" $x$ defined (up to an additive constant) by
\begin{equation} \label{tortoise-definition}
x(r) = \int \frac{dr}{f(r)} = r + r_H \log(1-r/r_H).
\end{equation}
Wave equations of the form \eqref{wave-equation} also characterize perturbations 
with $j=1,2$ \cite{wheeler,reggewheeler}; only the form of $V(r)$ changes.\footnote{For 
$j=2$ there is an added complication because one can consider
either ``axial'' or ``polar'' perturbations; \eqref{schwarzschild-4d-potential} strictly
applies only to the axial case, but the two are closely related and have the same
quasinormal frequencies \cite{Chandrasekhar:1985kt}.}  The 
three cases can be nicely summarized,
\begin{equation}  \label{schwarzschild-4d-potential}
V(r) = f(r) \left( \frac{l(l+1)}{r^2} + \frac{(1-j^2)r_H}{r^3} \right).
\end{equation}
In our computation it will be convenient to leave $j$ generic and then take the limit
$j \to 0$ or $j \to 2$ in the answer.  We note that our analysis will depend on the
$1/r^4$ term in $V(r)$ being dominant; it is therefore singular as $j \to 1$ and we cannot
predict the behavior there.

\subsection{Transmission and reflection in $d=4$ Schwarzschild} \label{section-transmission-schwarzschild-4d}

In this section we will compute the transmission and reflection coefficients for \eqref{wave-equation}, in the 
limit of large imaginary frequencies.  Most of the ideas required to do this 
calculation have appeared already in \cite{Motl:2003cd}, in particular the idea of using the boundary condition
at $r_H$ to provide the monodromy.  The general method of analytic continuation to complex $r$ is much older and 
has been used successfully for numerical calculations in the past (see \cite{glampand} for a recent 
discussion and review.)

\sfigc{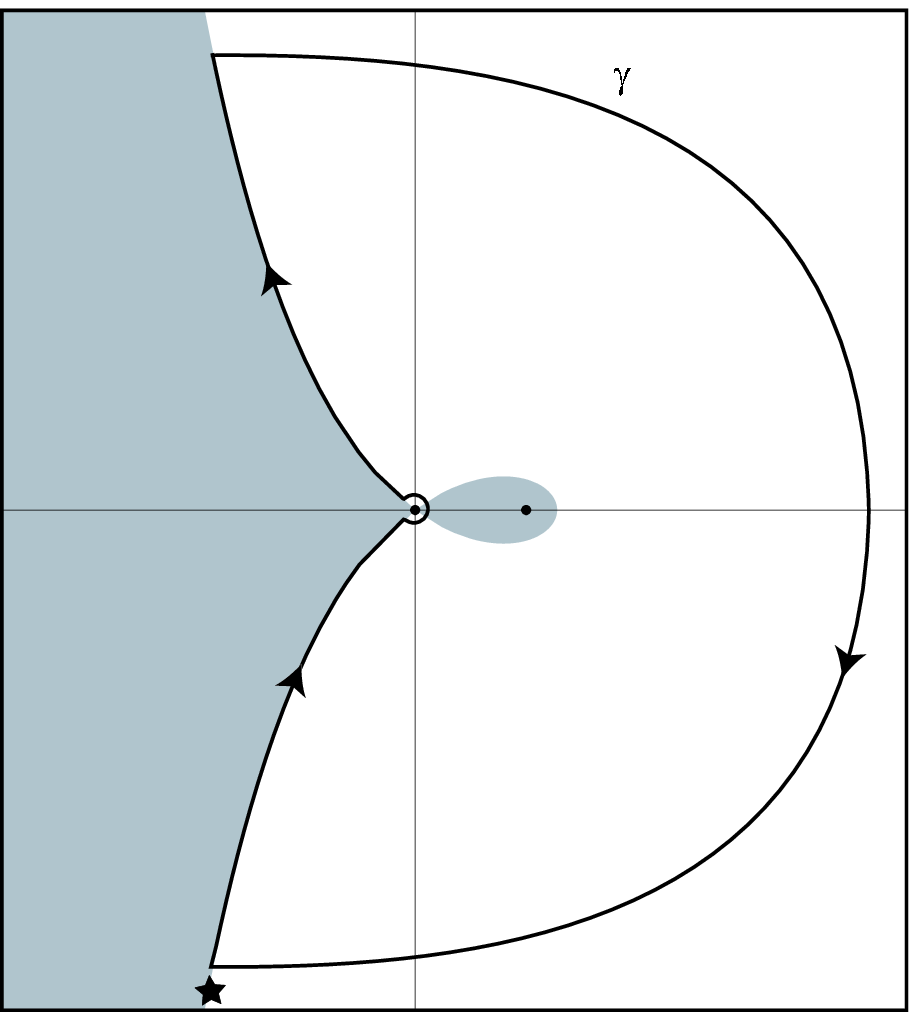}{The complex $r$-plane.  The $\re(x) < 0$ region 
is colored gray, and the points $r=0$, $r = r_H$ and a contour $\gamma$ for
calculation of the transmission and reflection coefficients are marked.  $\gamma$ runs along the line
$\re(x) = 0$ to $\abs{\omega x} \gg 1$ in each direction.  The symbol \spt represents the direction in which
the boundary conditions ``at infinity'' are imposed.}

Although we are studying a wave equation defined on the physical region $r_H < r < \infty$, to get the analytic 
continuation to imaginary $\omega$ we have to consider complex $r$.  The behavior of the tortoise
coordinate \eqref{tortoise-definition}
at complex $r$ plays an important role in the computation; in particular, since $x(r)$ contains
$r_H \log(1-r/r_H)$, it is multivalued, with the branches 
differing by $\Delta x = 2 \pi i n r_H$.  This multivaluedness is crucial for
our computation.  $\re x$, however, is 
uniquely defined; in Figure \ref{r-plane} we have indicated its sign by shading.

The idea of the computation is to study the behavior of a solution
$\psi(r)$ to \eqref{wave-equation} as we travel around a specific
contour $\gamma$, shown in Figure \ref{r-plane}.  We will calculate
the effect of the trip around $\gamma$ in two different ways:  a ``local'' computation by 
integrating the differential equation directly and a ``global'' computation using the boundary
conditions at the horizon.  Comparing the two gives enough information to calculate the 
transmission and reflection coefficients.

Since $V(r)$ vanishes faster than $1/\abs{r}$ as $r \to \infty$, 
as we run to infinity in any direction the asymptotics of a solution
to \eqref{wave-equation} are of the form
\begin{equation} \label{plane-wave-asymptotics}
\psi(r) \sim C_+ e^{+ i \omega x} + C_- e^{-i \omega x}.
\end{equation}
In \eqref{plane-wave-asymptotics} 
there is an ambiguity in defining $e^{\pm i \omega x}$, which arises from the multivaluedness of $x(r)$.
This ambiguity implies a compensating ambiguity in defining $C_\pm$, even for a fixed $\psi(r)$.  This ambiguity is compounded
by the fact that $\psi(r)$ itself can have multiple branches, since it is the solution to a differential
equation which has singularities.  The upshot is that as we travel 
around $\gamma$
and return to our starting point the coefficients $C_\pm$ do not return to their original values 
(i.e. they have a nontrivial monodromy), because we end up 
on a different branch both for $x(r)$ and for $\psi(r)$.\footnote{Actually, our boundary conditions
at $r_H$ will imply that only $C_-$ has nontrivial monodromy.}

As explained in Appendix \ref{appendix-generalities}, there is a branch cut in $T(\omega)$ and $R(\omega)$ on the positive imaginary $\omega$-axis.
If we are on the right side of the branch cut (taking $\re \omega$ infinitesimally positive) 
then $T(\omega)$ and $R(\omega)$ are defined by the boundary conditions
\begin{equation}
\psi \sim
\begin{cases}
e^{i \omega x} + R(\omega) e^{-i \omega x} & \as r \to \spt, \label{imaginary-transmission-boundary} \\
T(\omega) e^{i \omega x} & \as r \to r_H,
\end{cases}
\end{equation}
where $\psi(r)$ is a solution of \eqref{wave-equation}.  As
was done in \cite{Motl:2003cd}, we will obtain our results by computing
in two different ways the monodromy of the coefficient of $e^{-i \omega x}$.

To make this last statement clearer, let us study for a moment the behavior of $\psi(r)$ if we
simplify by setting $V(r)=0$.  In this case we may replace the asymptotic relation in 
\eqref{imaginary-transmission-boundary} by exact equality:
$\psi(r) = e^{i \omega x} + R(\omega) e^{-i \omega x}$ near \spt.  Then we follow $\psi(r)$ along 
$\gamma$.  Since locally we always have free wave propagation in the variable $x$, when
we return to \spt we still have $\psi(r) = e^{i \omega x} + R(\omega) e^{-i \omega x}$, 
i.e. the coefficient of $e^{-i \omega x}$ is the same before and
after the trip around $\gamma$.  On the other hand, $\gamma$ encloses a singularity 
$r=r_H$ around which both $x(r)$ and $\psi(r)$
are multivalued.  The boundary condition \eqref{imaginary-transmission-boundary} 
implies that $\psi(r)$ picks up a phase
$e^{2 \pi \omega r_H}$ on any path encircling 
$r_H$ in the clockwise direction.  Furthermore, $e^{-i \omega x}$ is multiplied by the phase
$e^{-2 \pi \omega r_H}$ on such a path.  These two
phases together imply that the coefficient of $e^{-i \omega x}$ 
must be multiplied by $e^{4 \pi \omega r_H}$ after the trip around $\gamma$.
But we already showed that it is unchanged after this 
trip; so we get $R(\omega) = e^{4 \pi \omega r_H} R(\omega)$, which implies $R(\omega)=0$.  
This is as expected:  there is no reflection, since 
there is no potential!  
We can calculate $T(\omega)$ by traveling on any contour 
running from \spt to the horizon; then, since we have free wave propagation, the boundary 
conditions at \spt given by \eqref{imaginary-transmission-boundary} are propagated directly 
to the horizon, and comparing with the boundary condition there we see immediately 
that $T(\omega)=1$.  
(We also see immediately that $R(\omega)=0$, which we had already concluded above by more 
elaborate means; the reason for the more complicated discussion above is that it generalizes 
to the case of a nontrivial potential.)

Now let us consider the real problem at hand, where $V(r)$ is 
given by \eqref{schwarzschild-4d-potential}:
\begin{equation} \label{schwarzschild-4d-potential-repeated}
V(r) = f(r) \left( \frac{l(l+1)}{r^2} + \frac{(1-j^2)r_H}{r^3} \right).
\end{equation}
The arguments above still show that $R$ is multiplied by the monodromy
$e^{4 \pi \omega r_H}$ after a trip around
$\gamma$; this depended only on the boundary condition \eqref{imaginary-transmission-boundary} at $r_H$
and the form of $x(r)$.  
Note that in the black hole context 
$4 \pi r_H$ is the inverse Hawking temperature $\bH = 1 / \THa$, 
so the monodromy can be written $e^{\bH \omega}$.

Now we follow $\gamma$ beginning at \spt where we have the boundary 
condition \eqref{imaginary-transmission-boundary}.  In 
Appendix A we describe a natural basis $\psi_\pm(r)$ of solutions near $r=0$;
as discussed there, by asymptotic matching
along the line $\re x = 0$ we can relate the form of $\psi$ slightly southwest of 
$r=0$ to the form of $\psi$ at \spt.  Namely, if we write the solution 
southwest of $r=0$ as $\psi(r) = A_+ \psi_+(r) + A_- \psi_-(r)$, and introduce the notation
\begin{equation} \label{notation-definition}
\n{m} = A_+ e^{i m (1+j)\pi/4} + A_- e^{i m (1-j)\pi/4},
\end{equation}
then matching the asymptotics \eqref{solution-asymptotics} of $\psi_\pm$ 
to \eqref{imaginary-transmission-boundary} gives
\begin{align}
\n{-1} & \approx 1, \label{asymptotic-relation-1} \\
\n{1} & \approx R(\omega). \label{asymptotic-relation-2}
\end{align}
Here $\approx$ means there are \err corrections.
Next we want to express $T(\omega)$ in terms of $A_\pm$.  For this we
first make a quarter-turn from southwest to southeast, then travel out along $\re x = 0$ 
until $\abs{\omega x} \gg 1$, and then run directly to $r=r_H$.
Since on this trip $\re x < 0$ and the $\omega^2$ term dominates $V(r)$, 
the WKB (physical optics) approximation guarantees that the coefficient of $e^{i \omega x}$
will not be modified as we travel to the horizon, to leading order in $\omega$.\footnote{We are
being slightly loose here, but the idea is that $e^{i \omega x}$
dominates the $\omega \to i \infty$ asymptotics when $\re x < 0$, so any modification to its coefficient
would cause those asymptotics to deviate from the
physical optics approximation.  This is the same idea used in the old derivations of the WKB transmission formula
for a particle with energy near the top of a barrier \cite{Kemble}; we will use it several times.}  
Therefore this coefficient at the horizon is determined
simply by the asymptotics of $\psi$ as we run southeast from $r=0$, which are given by
\eqref{solution-rotation} and \eqref{solution-asymptotics}, leading to
\begin{equation}
\n{3} \approx T(\omega). \label{asymptotic-relation-3}
\end{equation}
Finally we want to complete our trip around the contour.  So we make a three-quarter turn from
southwest to northwest, then travel north until $\abs{\omega x} \gg 1$.  Then the asymptotic matching
using \eqref{solution-rotation}, \eqref{solution-asymptotics} determines the coefficient of $e^{-i \omega x}$ 
to be $\n{5}$.
On the other hand, we can follow this coefficient around the big semicircle and back to \spt; again
by the WKB approximation it must be unchanged by this trip.  But as discussed above, after the full
round trip
around $\gamma$ the coefficient of $e^{-i \omega x}$ is $e^{\beta \omega} R(\omega)$, so we get
\begin{equation}
\n{5} \approx e^{\beta \omega} R(\omega). \label{asymptotic-relation-4}
\end{equation}
The relations \eqref{asymptotic-relation-1}, \eqref{asymptotic-relation-2}, \eqref{asymptotic-relation-3},
\eqref{asymptotic-relation-4} 
define an inhomogeneous linear system for the four unknowns $A_+, A_-, T(\omega), R(\omega)$.  
Solving it gives the final answer:
\begin{align} 
T(\omega) & \approx \frac{e^{\beta \omega} - 1}{e^{\beta \omega} + (1 + 2 \cos \pi j)}, \label{transmission-schwarzschild-4d} \\
R(\omega) & \approx 2i \frac{\cos \half \pi j}{e^{\beta \omega} + (1 + 2 \cos \pi j)}. \label{reflection-schwarzschild-4d}
\end{align}
Here the symbol $\approx$ means there can be \err corrections both in the numerator and the denominator.
The previous analytical results of \cite{Motl:2003cd,
Motl:2002hd} for the asymptotic quasinormal frequencies are recovered by looking at the poles of $R(\omega)$ and $T(\omega)$, 
which occur at $\omega$ satisfying
\begin{equation} \label{asymptotic-qnf-schwarzschild-4d}
e^{\beta \omega} + (1 + 2 \cos \pi j) \approx 0.
\end{equation}
In particular, for $j=0$ or $2$ we recover $\re \omega \approx \THa \log 3$, which was important for \cite{Hod:1998vk,Dreyer:2002vy}.

The results \eqref{transmission-schwarzschild-4d}, \eqref{reflection-schwarzschild-4d} apply for $\omega$ approaching
the branch cut from the right, i.e. $\re \omega$ infinitesimally positive.  
If instead we took $\re \omega$ infinitesimally negative then the boundary condition at infinity in
\eqref{imaginary-transmission-boundary} would be set at the north end of Figure \ref{r-plane} rather than
at \spt, and $\gamma$ would run in the opposite direction.  The effect is to replace $e^{\bH \omega}$ by
$e^{-\bH \omega}$ in \eqref{asymptotic-relation-4} and hence to replace $\omega$ by $-\omega$ in the final result
(still valid for $\omega \to i \infty$.)
Then we recover for $j=0$ or $2$ the quasinormal frequencies on the other side of the imaginary axis:  $\re \omega \approx 
-\THa \log 3$.

We can also study $T(-\omega)$ and $R(-\omega)$ as $\omega \to i\infty$.  
The analysis is similar to that given above with two
differences.  First, there are some signs that have to be reversed in the asymptotics \eqref{solution-asymptotics}.  
Taking these signs into account, the analogues of \eqref{asymptotic-relation-1}, \eqref{asymptotic-relation-2}, 
\eqref{asymptotic-relation-3} are
\begin{align}
\n{1} & \approx 1, \\
\n{-1} & \approx R(-\omega), \\
\n{1} & \approx T(-\omega). \\
\intertext{Second, the WKB approximation
on the large semicircle is now good for the coefficient of $e^{i \omega x}$ rather than for $e^{-i \omega x}$.
For this coefficient the two monodromy factors $e^{2 \pi \omega r_H}$ cancel instead of multiplying, giving
instead of \eqref{asymptotic-relation-4}}
\n{5} & \approx 1. \\
\intertext{Solving, we obtain}
T(- \omega) & \approx 1, \\
R(- \omega) & \approx -2i \cos \half \pi j.
\end{align}
One can also see that $T(-\omega) \approx 1$ directly just by 
noting that (as discussed in Appendix \ref{appendix-generalities})
when $\im \omega > 0$ the analytic continuation is not necessary for defining $T(-\omega)$ and $R(-\omega)$;
we can just work directly on the real line, and then the WKB approximation gives
$T(-\omega) \approx 1$ immediately since there are no turning points.

\subsection{Transmission and reflection for other black holes} \label{section-other-black-holes}

The computation of the transmission and reflection coefficients given in Section 
\ref{section-transmission-schwarzschild-4d} generalizes in a straightforward way to the other black holes 
considered in \cite{Motl:2003cd}, with modifications in each case
just like the modifications made in \cite{Motl:2003cd} for the computation of the
quasinormal frequencies.  Here we only report the answers.  For Schwarzschild
in $d \ge 4$ coupled to a massless scalar\footnote{The restriction to scalar fields arises because
the appropriate generalization of \eqref{schwarzschild-4d-potential} is not known for arbitrary metric
perturbations in $d \ge 4$.} the result is just as it was in $d=4$:  as $\omega \to i \infty$,
\begin{align} 
T(\omega) & \approx \frac{e^{\beta \omega} - 1}{e^{\beta \omega} + 3}, \label{transmission-schwarzschild-repeated} \\
R(\omega) & \approx \frac{2i}{e^{\beta \omega} + 3}, \label{reflection-schwarzschild-repeated} \\
T(-\omega) & \approx 1, \label{transmission-schwarzschild-negative-repeated} \\
R(-\omega) & \approx -2i. \\
\intertext{For \rn in $d=4$ coupled to massless scalar or electromagnetic-gravitational waves the result is}
T(\omega) & \approx \frac{e^{\beta \omega} - 1}{e^{\beta \omega} + 2 + 3e^{-\bHI \omega}}, \label{transmission-rn-4d-repeated} \\
R(\omega) & \approx \pm i\sqrt{3} \frac{1 + e^{-\bHI \omega}}{e^{\beta \omega} + 2 + 3e^{-\bHI \omega}}, \\ \label{reflection-rn-4d-repeated}
T(-\omega) & \approx 1, \\
R(-\omega) & \approx \mp i \sqrt{3},
\end{align}
where the sign is determined by the type of field we consider.
A useful check on the results is provided by the analytically-continued conservation of flux,
\begin{equation}
T(\omega) T(-\omega) + R(\omega) R(-\omega) = 1.
\end{equation}

\subsection{Hawking radiation and greybody factors} \label{section-hawking}

Taken together, the results of Section \ref{section-other-black-holes}
for the transmission coefficient at large imaginary frequencies
have a rather suggestive structure.  To understand it, let us
first recall one physical role played by the transmission coefficient
at \ti{real} frequencies.

The spectrum of Hawking radiation
seen by a static observer at spatial infinity is not thermal; rather,
the particle flux obeys \cite{Hawking:1975sw}
\begin{equation} \label{hawking-formula}
F(\omega) \propto \frac{\abs{T(\omega)}^2}{e^{\beta \omega} - 1}.
\end{equation}
The interpretation of this formula is that Hawking
radiation is produced with a thermal spectrum at the horizon and then the spacetime curvature between the horizon and infinity scatters some of the radiation back down the hole, so that the thermal spectrum is multiplied by the ``greybody factor'' $\abs{T(\omega)}^2$.

In a remarkable paper \cite{Maldacena:1997ix} it was shown
that, for a certain class of near extremal five-dimensional black holes admitting a D-brane construction 
\cite{Strominger:1996sh}, 
and for energies small compared to $1/r_H$, the greybody factor allows semiclassical general relativity to
reproduce the emission spectrum of the appropriate system of D-branes.  Namely, for small $\omega$
and in a certain ``dilute gas'' regime of the black hole parameter space, a direct classical computation gives
\begin{equation} \label{stringy-transmission}
\abs{T(\omega)}^2 \approx \frac{e^{\bH \omega} - 1}{(e^{\bL \omega/2} - 1)(e^{\bR \omega/2} - 1)},
\end{equation}
Here $\bL$ and $\bR$ are parameters in the classical black hole solution, which
can be identified with inverse temperatures characterizing the left-moving and
right-moving excitations in an effective SCFT
which describes the D-brane degrees of freedom.  The factor $e^{\beta \omega} - 1$ in \eqref{stringy-transmission}
cancels the denominator of \eqref{hawking-formula}, leaving the spectrum
observed from infinity:
\begin{equation} \label{stringy-spectrum}
F(\omega) \propto \frac{1}{(e^{\bL \omega/2} - 1)(e^{\bR \omega/2} - 1)}.
\end{equation}
But \eqref{stringy-spectrum} is exactly
the emission spectrum calculated from the effective SCFT!  
To quote \cite{Maldacena:1997ix}, ``to the observer at infinity
the black hole, masquerading in its greybody cloak, looks like the string, for energies
small compared to the inverse Schwarzschild radius of the black hole.''
Of course, this result was
discovered after the D-brane construction was already known; but if the order had been reversed, one might 
have speculated that this emission spectrum gave a potential clue to the microscopic
description of the black hole.  This perspective was pursued further
in \cite{Maldacena:1997ih} where the calculation of the greybody factor was extended to near extremal 
charged rotating black holes, both in $d=4$ and $d=5$, and again
found to be consistent with an effective SCFT description.

In the present case of the Schwarzschild or \rn black hole we have a similar result, 
with the important caveat that it does not apply directly to the emission spectrum at real $\omega$.  
Rather, suppose an observer at spatial 
infinity measures the \ti{exact} emission spectrum (for a massless scalar field
minimally coupled to the curved background) over some interval of real 
$\omega$ and then analytically continues to large imaginary $\omega$.  What is the result?  Using the fact 
that $T(\omega)^* = T(-\omega)$ for real $\omega$, together
with \eqref{transmission-schwarzschild-repeated} and \eqref{transmission-schwarzschild-negative-repeated}, 
the analytic continuation of $\abs{T(\omega)}^2$ to large imaginary $\omega$ is 
\begin{equation}
T(\omega) T(-\omega) \approx \frac{e^{\beta \omega} - 1}{e^{\beta \omega} + 3}.
\end{equation}
So substituting in \eqref{hawking-formula} we find the same cancellation 
between the numerator and denominator that occurred in \cite{Maldacena:1997ix};
the observer at infinity would compute a
flux which, for large imaginary $\omega$, depends on $\omega$ as
\begin{equation} \label{analytically-continued-spectrum-schwarzschild}
F(\omega) \propto \frac{1}{e^{\beta \omega} + 3}.
\end{equation}
Similarly, in the case of the \rn black hole the result is
\begin{equation} \label{analytically-continued-spectrum-rn-4d}
F(\omega) \propto \frac{1}{e^{\beta \omega} + 2 + 3 e^{-\bHI \omega}},
\end{equation}
where $\bHI$ is the inner horizon temperature.

What could these formulas mean?  The fact that they are always written
in terms of Boltzmann factors (or more properly ``Boltzmann phases'' since $\omega$ is
pure imaginary) is suggestive and we propose that, 
just as the spectrum of Hawking radiation of a near extremal black hole
at \ti{small real} $\omega$ can be obtained from an effective SCFT, the spectrum of
a Schwarzschild or \rn black hole at
\ti{large imaginary} $\omega$ also admits an interpretation in terms of some new
degrees of freedom.  These degrees of freedom would be expected to 
have rather exotic statistics (the 
possibility that the quasinormal modes of $d=4$ Schwarzschild could imply 
``tripled Pauli statistics'' appeared also in \cite{Motl:2002hd}.)
In the \rn case the occurrence of both $e^{\bH \omega}$ and $e^{\bHI \omega}$ suggests further that 
they should live on both the inner and outer horizons (such descriptions have been considered
before, e.g. in \cite{Vaz:2000mq}.)  At any rate, the bulk scattering amplitudes we have computed could
then be related to correlation functions in this conjectural ``boundary'' theory.  Since our result
is valid around large imaginary frequencies it is natural to suspect that this boundary
theory bears some relation to the Euclidean continuation of the black hole.

One might wonder why the structure we observe was not already visible in
\cite{Maldacena:1997ih}.  After all, that paper computed precisely the same function we are
computing.  The reason our $T(\omega) T(-\omega)$ is not simply the analytic continuation
of the $\abs{T(\omega)}^2$ in \cite{Maldacena:1997ih} is that both are approximations to 
the true answer, valid in different
regimes.  If we know a function $f(\omega)$ exactly at real $\omega$, then we can analytically continue
it to get the exact answer for imaginary $\omega$; but if we neglect a small correction at real $\omega$
it can invalidate the continuation to imaginary $\omega$.  This problem
also occurs when using AdS/CFT to obtain 
Minkowski space boundary correlators, as discussed recently in \cite{Son:2002sd,Herzog:2002pc}.

\section{Comparison to numerical results} \label{section-numerical-results}

Since some of the asymptotic formulas computed here and in \cite{Motl:2003cd,Motl:2002hd}
are still new and the result for \rn is rather strange, it may be useful to summarize the numerical 
evidence supporting them.  While I am not aware of any numerical results on the 
full asymptotic transmission coefficient, there have been several studies of 
the asymptotic quasinormal frequencies for various types of black hole, to which the analytical results
may usefully be compared.

In the case of the Schwarzschild black hole the only numerical results
are in $d=4$ and they support the conclusion that the asymptotic real part of the frequency
is $\THa \log 3$, at least for $j=2$ perturbations \cite{leaverkerr, nollertqnm, andersson, bachelot}.
I do not know of any literature on the asymptotic quasinormal frequencies for other $j$.  In
\cite{nollertqnm} the corrections at finite $\omega$ were also tabulated; the results there are
consistent with the hypothesis that the
first subleading correction is \err and proportional to
$(l^2+l-1)$.\footnote{This proportionality was discovered by Martin Schnabl.}  This is consistent
with an estimate of the correction to the Bessel function asymptotics 
as discussed in Appendix \ref{appendix-asymptotics}, but I have not managed to evaluate the exact
coefficient to compare with \cite{nollertqnm}.  Recently \cite{MaassenvandenBrink:2003as} appeared
containing a computation closely related to the one given here, 
obtaining the discontinuity across the branch cut of the 
solution satisfying the outgoing boundary condition at infinity as well as the first 
correction; this correction was found to be proportional to $(l^2+l-1)$ and it is likely that those results
can be used to get the correction to the transmission and reflection coefficients as well.

In the case of the \rn black hole in $d=4$, the asymptotic formula for the quasinormal frequencies
is more intricate.
It is convenient to introduce the parameter
\begin{equation} \label{new-param-definition}
\alpha = \frac{1}{1 - (r_I/r_H)^2},
\end{equation}
where $r_I$ and $r_H$ are the inner and outer horizon radii respectively; $\alpha$ runs from $1$ for the
uncharged black hole (Schwarzschild limit) to $\infty$ for the extremal black hole.  In terms of $\alpha$
and $\oh = 8 \pi G M \omega$, the requirement that the transmission coefficient \eqref{transmission-rn-4d} 
has a pole becomes
\begin{equation} \label{asymptotic-qnf-reissner-nordstrom-4d-rewritten}
e^{\alpha \oh} + 3 e^{(\alpha-1) \oh} + 2 \approx 0.
\end{equation}
The behavior in the asymptotic limit depends crucially on whether $\alpha$ is rational.  
If $\alpha = p/q$ then we can write $x = e^{\oh / q}$ and \eqref{asymptotic-qnf-reissner-nordstrom-4d-rewritten}
becomes a polynomial equation in $x$, 
with finitely many solutions.  After taking the logarithm each solution then gives an infinite 
family of quasinormal frequencies, with constant real part and evenly spaced imaginary part; 
so instead of just one
asymptotic real part we have a finite number of them.  For example, if $\alpha=2$ we can set $x = e^{\oh}$ 
and then get $x^2 + 3x + 2 = 0$, which has two solutions $x=-1$, $x=-2$, giving two families of asymptotic
quasinormal frequencies, at $\oh = 2\pi i(n+\half)$ and $\oh = 2\pi i(n+\half) + \log 2$.  On the other hand, if
$\alpha$ is irrational, the behavior of the asymptotic quasinormal frequencies is more complicated; it
probably admits some statistical description in terms of the continued fraction representation of $\alpha$.
Such a sensitive dependence on whether $\alpha$ is rational seems strange from the point of view 
of general relativity, but it might not be so strange from the point of view of an underlying microscopic 
description of the black hole which makes the parameters discrete.

We can also follow the behavior of
a single quasinormal frequency as we increase $\alpha$ from $1$ to $\infty$.  At $\alpha=1$ the solutions
of \eqref{asymptotic-qnf-reissner-nordstrom-4d-rewritten} are simply $\oh = 2\pi i(n+\half) + \log 5$.\footnote{The
reader may be alarmed that $\alpha=1$ is the Schwarzschild limit and we are getting $\re \omega \approx 
\THa \log 5$ instead 
of the correct $\re \omega \approx \THa \log 3$ for Schwarzschild.  Such a reader is 
urged to suspend disbelief for a few more paragraphs.}  Increasing $\alpha$
continuously the frequency traces a spiral in the complex plane, shown in Figure \ref{spiral-analytical-5} for $n = 5$.
\fig{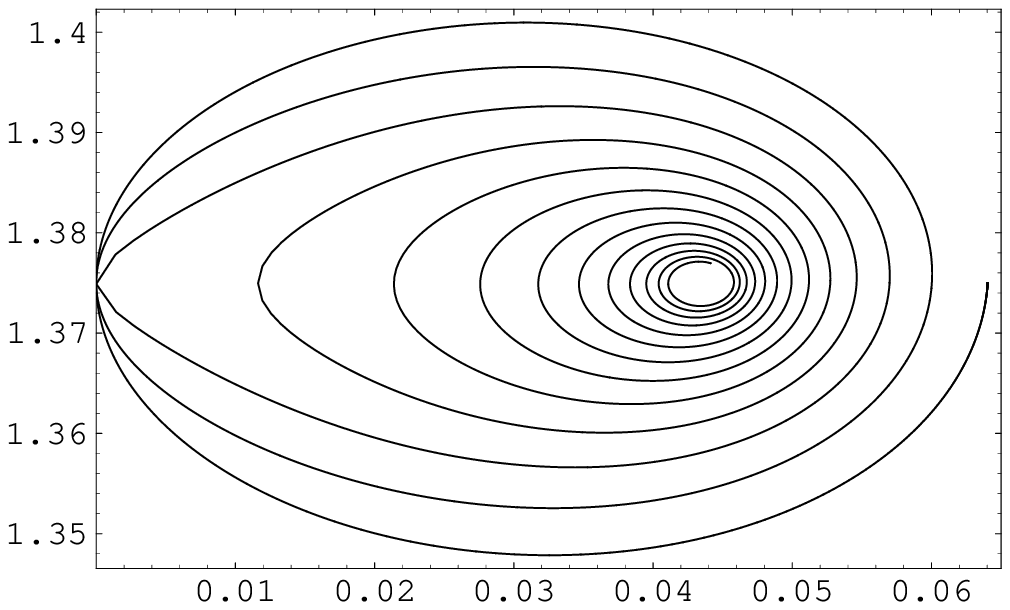}{Inspiraling behavior of the asymptotic formula 
\eqref{asymptotic-qnf-reissner-nordstrom-4d-rewritten} for the $n=5$ quasinormal frequency as the charge is 
increased toward extremality.  The axes are $\re GM \omega$ and $\im GM \omega$.  The center of the spiral is
at $GM\omega = (11/8)i + \log 3 / 8\pi$.}
The center of the spiral corresponds to the extremal limit $\alpha \to \infty$ in which we may neglect 
the last term of \eqref{asymptotic-qnf-reissner-nordstrom-4d-rewritten}, obtaining simply
$e^{\oh} + 3 \approx 0$, which implies 
\begin{equation}
8 \pi GM \omega \approx 2\pi i \left(n+\half \right) + \log 3.
\end{equation} 
Amusingly, this result in the extremal limit agrees with the asymptotic quasinormal 
frequency for a Schwarzschild black hole of the same mass.

The spiral structure shown in Figure \ref{spiral-analytical-5} was visible already in the low-order modes computed
in \cite{Andersson:1996xw}.
But very recently the full asymptotic formula has been confirmed 
in detail by the numerical analysis of \cite{Berti:2003zu}, and looking at their results
is useful not only to convince oneself that the formula is asymptotically correct but also
to understand the nature of the \err corrections.
Rather than looking at the spiral it is convenient to
plot $\re \omega$ and $\im \omega$ separately as functions of the
charge $Q$; this is done in Figure \ref{figure-comparison-array}
for $n = 30, 5000, 10000$.\footnote{I am indebted to E. Berti and K. Kokkotas for sharing their data files with me, and to L. Motl for preparing earlier versions of these figures as well as suggesting that they be made in the first place.}

\newcommand{\comparisonblock}[1]{
\unitlength1in
\begin{minipage}[t]{3.2in}
\begin{picture}(3.2,1.95)\includegraphics[width=3.2in]{figures/real-comparison-#1.eps}\end{picture}\par 
\end{minipage}
\hfill
\begin{minipage}[t]{3.2in}
\begin{picture}(3.2,1.95)\includegraphics[width=3.2in]{figures/imag-comparison-#1.eps}\end{picture}\par
\end{minipage}}

\begin{figure} \label{figure-comparison-array}
\comparisonblock{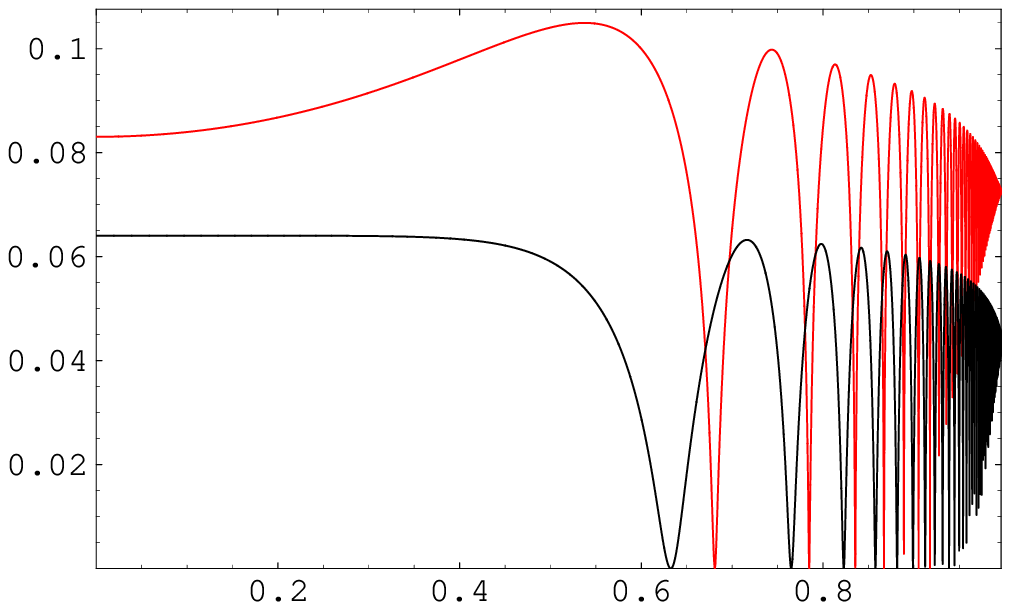}
\comparisonblock{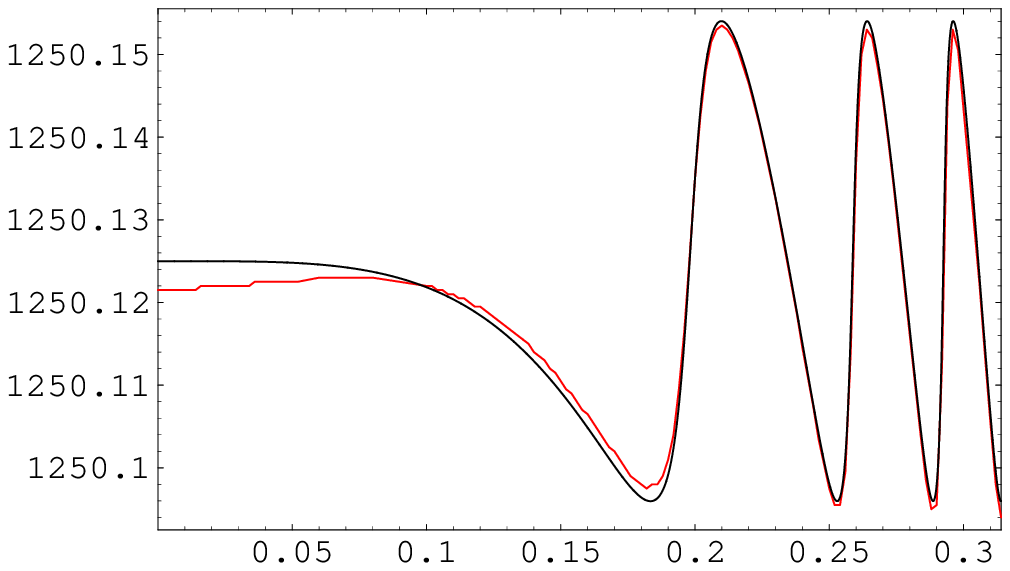}
\comparisonblock{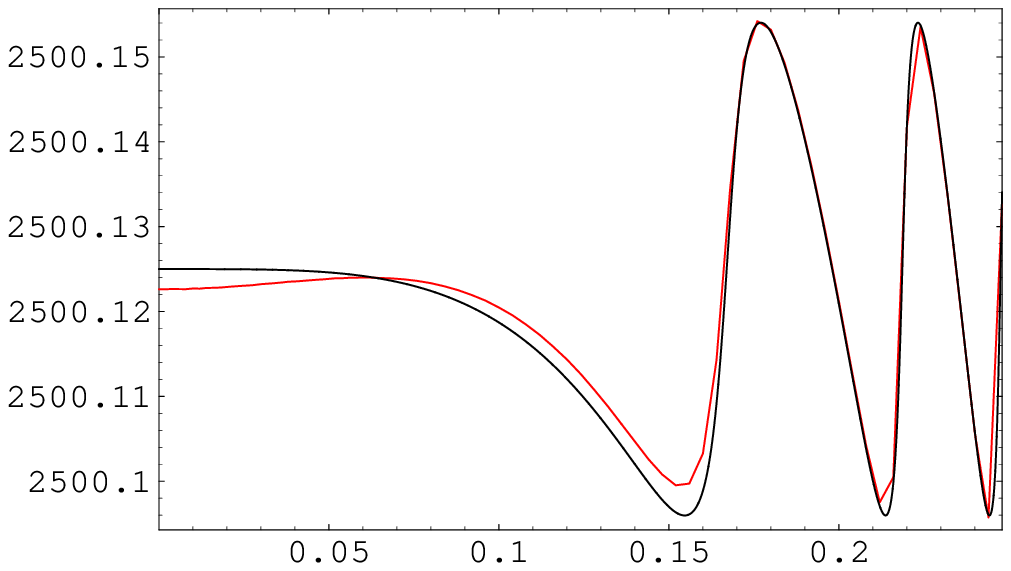}
\caption{Real parts (left) and imaginary parts (right) of quasinormal frequencies for the \rn black hole,
at mode number (top to bottom) $n=30,5000,10000$.  The red line is the numerical data of
\cite{Berti:2003zu} and the black line is the asymptotic formula \eqref{asymptotic-qnf-reissner-nordstrom-4d-rewritten}.  
The vertical axis is (Re or Im) $GM \omega$; the horizontal is $Q$, in units where $Q=1$ is extremal.}
\end{figure}

We see that at large $n$ the numerical and analytical results agree extremely well, except at small $Q$
where the real part is significantly different even at $n=10000$.  How is such a disagreement possible
in the asymptotic limit?  The point is that, 
as already mentioned in \cite{Motl:2003cd}, for the asymptotic formula to be valid one has to take the $n \to \infty$ limit at fixed $Q \neq 0$.  In this limit we will eventually find agreement with the asymptotic formula
(e.g. if we fix $Q = 0.16$, we can see from Figure \ref{figure-comparison-array} 
that $n=5000$ is not sufficiently large for
the asymptotic behavior to take over, but $n=10000$ is.)  The required $n$ diverges as $Q \to 0$, or
put another way, the \err corrections to the asymptotic formula \eqref{asymptotic-qnf-reissner-nordstrom-4d-rewritten}
appear with a coefficient which diverges as $Q \to 0$.  A more systematic analysis of these corrections should 
demonstrate this behavior explicitly; here we only remark that the analysis in \cite{Motl:2003cd} which led to 
\eqref{asymptotic-qnf-reissner-nordstrom-4d-rewritten} is certainly singular in the limit $Q \to 0$, so we would 
expect that the corrections would indeed blow up.

\section{Conclusions}

In this paper we have presented some results about the scattering of free massless quantum fields by
a black hole, in the limit of large imaginary frequencies.  These results stand independent of 
any speculation about their meaning, but their form is tantalizing and it seems that they deserve 
to have a simple explanation.  The remarks about a possible boundary interpretation 
which appear in Section \ref{section-hawking} are of course pure conjecture at the moment;
hopefully the situation will be clarified in the future.

It remains unclear whether
one can connect the computation presented here to the area-quantization ideas 
which motivated the consideration of 
the high-imaginary-frequency quasinormal modes in \cite{Hod:1998vk,Dreyer:2002vy}.
It is clear that these ideas will have to be modified or reinterpreted in some way if they 
are to be correct, since the \rn and Kerr black holes do not share the simple behavior
seen in the Schwarzschild case.  
Nevertheless, it is possible that this could be done, and
the spin-network picture discussed in \cite{Dreyer:2002vy} could be a candidate for
the boundary description we are conjecturing.

There are several other calculations along the same lines which could be done and might shed some light
on the situation.  One possibility would be to include chemical potentials, either by studying a rotating
black hole or a charged scalar in the \rn background.  In both cases one faces the technical difficulty that the
potentials appear to be long-range (they involve $1/r$) and hence the behavior is not simply plane-wave
at infinity.  In the rotating case, at least for the Kerr black hole, 
this problem can be avoided by a clever change of variable introduced by Sasaki
and Nakamura \cite{Sasaki:1982kj},\footnote{I thank Shinji Mukohyama and Eanna Flanagan 
for informing me about this reference.} but the problem remains difficult because the angular equation
also has to be treated approximately (the centrifugal potential, which was just $l(l+1) / r^2$ for Schwarzschild
and had no effect to leading order in $\omega$, depends
on $\omega$ for Kerr.)
For \rn I am not aware of a change of variables which makes the
potential short-range, but it should be possible to make one and this problem might be more tractable
than Kerr.  On the other hand, in the Kerr 
case there are some numerical results available \cite{Berti:2003zu}
which look puzzling but might be helpful.  One could also study the high-imaginary-frequency behavior
for fermionic fields.  Here the relevant radial equation has been given e.g. in 
\cite{Mukhopadhyay:1999gf}, and the low overtone quasinormal modes were recently considered
in \cite{Cho:2003qe}.

Another possibility would be to apply our method to
the five-dimensional near extremal 
black holes studied in \cite{Maldacena:1997ix,Strominger:1996sh}, 
for which we already
have not only an effective SCFT but a full microscopic description, 
and see whether the resulting large-imaginary-frequency behavior can fit
into that description somewhere.

\section*{Acknowledgements}

I thank N. Arkani-Hamed, A. Ashtekar, E. Berti, O. Dreyer, E. Flanagan, C. Herzog, K. Kokkotas, K. Krasnov, 
A. Maassen van den Brink, A. Maloney, 
S. Minwalla, S. Mukohyama, H. Nastase, M. Schnabl, 
M. Schwartz, D. Son, A. Starinets, A. Strominger and C. Vafa for discussions.  I especially want to 
thank Lubo\v{s} Motl for numerous discussions and continuous prodding, as well as for a continuing stimulating 
collaboration.  I also thank California Institute of Technology for kind hospitality while this work was being
finished.  My research is supported by an NDSEG Graduate Fellowship.

\appendix
\section{Generalities on one-dimensional wave equations} \label{appendix-generalities}

The notions of transmission and reflection coefficient for a one-dimensional wave equation are well known, at
least for real frequencies.  However, there are some subtleties associated with analytic 
continuation to complex frequencies.  Under fairly general conditions this analytic continuation does
exist; the questions are what its singularities are and whether we can construct it in a way which facilitates
explicit computation.

Suppose given a wave equation on the real line $-\infty < x < \infty$
with the form of \eqref{wave-equation},
\begin{equation} \label{wave-equation-repeated}
\left[ - \dddx + V(x) - \omega^2 \right] \psi(x) = 0,
\end{equation}
where $V(x) \to 0$ faster than $1/\abs{x}$ as $x \to \pm \infty$.
We can then define $T(\omega), R(\omega)$ for waves of real frequency $\omega \neq 0$
by fixing the boundary conditions 
\begin{equation}
\psi(x) \sim
\begin{cases}
T(\omega) e^{+i \omega x}  \label{transmission-boundary} & \as x \to -\infty, \\
e^{+i \omega x} + R(\omega) e^{-i \omega x} & \as x \to +\infty. 
\end{cases}
\end{equation}

The boundary conditions 
\eqref{transmission-boundary} also suffice to extend
$T(\omega)$ and $R(\omega)$ to 
complex-analytic functions on the lower half-plane $\im \omega < 0$.  Furthermore these functions
are nonsingular; there cannot be any poles because they would imply the existence 
of a normalizable solution to \eqref{wave-equation-repeated}
with complex $\omega$, which is impossible because of the self-adjointness of $- d^2/dx^2 + V(x)$.\footnote{This 
self-adjointness fails for non-normalizable $\psi(x)$, which formally explains the existence of quasinormal modes,
i.e. poles of $T(\omega)$ and $R(\omega)$ in the upper half-plane.}

It is not straightforward to define
$T(\omega)$ or $R(\omega)$ 
for $\omega$ in the upper half-plane, because in that case the boundary condition 
\eqref{transmission-boundary} as $x \to +\infty$ 
does not give a constraint on the solution; the solution $e^{+i \omega x}$ 
is exponentially small compared to $e^{-i \omega x}$ as $x \to +\infty$, so adding $e^{+i \omega x}$ 
with an arbitrary coefficient does not affect the asymptotics as $x \to +\infty$, and there is no way to
define what we mean by requiring this coefficient to be $1$.  In some cases we
can nevertheless analytically continue $T(\omega)$ to the upper half-plane; 
the existence and singularity structure of such a
continuation depend on the properties of $V(x)$.  We now turn to some examples.

The simplest possibility is $V(x) = 0$ for all $x$.  In this case
the prescription above gives simply $T(\omega) = 1, R(\omega) = 0$ for 
all $\omega$ in the lower half-plane, which has the obvious analytic 
continuation $T(\omega) = 1, R(\omega) = 0$ for all $\omega$.

Another possibility is that $V(x) \neq 0$ but $V(x)$ strictly vanishes as $x \to \pm \infty$ (e.g. a finite 
square well).  In that case, for any $\omega$, the solutions of \eqref{wave-equation-repeated} near $\pm \infty$ 
are strict 
linear combinations of $e^{\pm i \omega x}$.  Therefore $T(\omega)$ can be uniquely defined for any $\omega$ 
just by replacing the asymptotic relation $\sim$ by exact equality in \eqref{transmission-boundary}.  
This prescription defines $T(\omega)$ and $R(\omega)$ univalently for any $\omega \neq 0$, so $T(\omega)$ 
and $R(\omega)$ are single-valued, although they can have isolated singularities at $\omega = 0$ and in 
the upper half-plane.

A third possibility is that $V(x)$ admits analytic continuation to complex $x$ and furthermore 
$V(x) \to 0$ faster than $1 / \abs{x}$ as $x \to \infty$ in any direction (e.g. $V(x) = 1/(x^2+1)$.)  
Note that this case and the one
just considered are mutually exclusive.  In this case we can define $T(\omega)$ by
setting our boundary conditions on the line $\omega x \in \R$ rather than $x \in \R$:
\begin{equation}
\psi(x) \sim \begin{cases}
T(\omega) e^{+i \omega x}  \label{transmission-boundary-complex} & \as \omega x \to -\infty, \\
e^{+i \omega x} + R(\omega) e^{-i \omega x} & \as \omega x \to +\infty. 
\end{cases}
\end{equation}
The advantage of \eqref{transmission-boundary-complex} over \eqref{transmission-boundary}
is that the solutions are oscillatory on the line $\omega x \in \R$ and hence the asymptotics \eqref{transmission-boundary-complex} determine $T(\omega)$ and $R(\omega)$.  
There is one additional complication:  if $V(x)$ has singularities along the line 
$\omega x \in \R$, then $\psi(x)$ will generally be 
multivalued, so we have to specify the contour along which we travel from one side of the line $\omega x \in \R$
to the other.  To make $T(\omega)$ and $R(\omega)$ holomorphic in $\omega$ the only option is to continuously
deform the contour by ``Wick rotation'' from the real axis; but then after a full $2 \pi$ rotation
the contour will hang on the singularities, so $T(\omega)$ and $R(\omega)$ are potentially 
multivalued.

Finally we come to the case which actually occurs for black hole transmission coefficients, say for $d=4$ Schwarzschild.
Here the analytic structure of $T(\omega)$ and $R(\omega)$ has been extensively studied because it is relevant for
the long time behavior of black hole perturbations; see in particular \cite{leaverbranchcut}.
In this case $V(x) = V[r(x)]$, with $r(x)$ defined implicitly by \eqref{tortoise-definition},
and while $V(r)$ is manifestly single-valued all over the complex $r$-plane,
$r(x)$ generally has multiple branches (because of the logarithm in \eqref{tortoise-definition}).
  So the analytic continuation of $V(x)$ is a multivalued 
function of $x$.  Nevertheless, since the only singularity of $V(r)$ occurs 
at $r=0$, the singularities of $V[r(x)]$ can only occur at
$x$ for which some branch of $r(x)$ has $r(x) = 0$, namely at $x = 2\pi i n$.  So $T(\omega)$, $R(\omega)$ can be
defined by Wick rotation of the $x$-plane contour to any $\omega$ with $\re \omega \neq 0$.  However, at $\re
\omega = 0$ we encounter an infinite chain of singularities and we have to choose which way to go around them;
our choice will depend on which way we made the Wick rotation (i.e. whether we are coming from $\re \omega > 0$ 
or $\re \omega < 0$).  $T(\omega), R(\omega)$ therefore have branch cuts at positive imaginary $\omega$.
In the body of the paper we compute these functions for $\omega$ on the right side of the cut, 
i.e. $\re \omega$ infinitesimally positive; in that case, after performing the Wick rotation in the $x$-plane, 
\eqref{transmission-boundary-complex}
corresponds to (see Figure \ref{r-plane} in Section \ref{section-transmission-schwarzschild-4d})
\begin{equation}
\psi(r) \sim
\begin{cases}
T(\omega) e^{+i \omega x} & \as r \to r_H, \\
e^{i \omega x} + R(\omega) e^{-i \omega x} & \as r \to \spt. \label{imaginary-transmission-boundary-repeated}
\end{cases}
\end{equation}
If we had assumed $\re \omega < 0$ instead, we would have put \spt at the north of Figure \ref{r-plane} instead
of the south.

\section{Asymptotics of solutions to the wave equation} \label{appendix-asymptotics}

In our analysis of the solutions of \eqref{wave-equation} along the line $\re x = 0$ we need to 
know their asymptotics in the limit $\omega \to i \infty$.  In this limit the term $\omega^2$ dominates
the potential everywhere except 
near $r=0$, so we expect that the interesting behavior will occur there.
Near $r=0$ we have from \eqref{tortoise-definition}
\begin{equation}
x/r_H = r/r_H + \log (1-r/r_H) = -\half(r/r_H)^2 - \frac{1}{3}(r/r_H)^3 + {\mathcal O}\left((r/r_H)^4\right)
\end{equation}
which can be inverted to obtain
\begin{equation}
r/r_H = \pm i \sqrt{2x/r_H} + 2x/3r_H + {\mathcal O}\left((x/r_H)^{3/2}\right)
\end{equation}
Substituting this in \eqref{wave-equation}, \eqref{schwarzschild-4d-potential} we obtain
the leading-order behavior of \eqref{wave-equation} near the $r=0$ singularity together
with its first correction:
\begin{equation} \label{wave-equation-singularity-corrected}
\left[ -\dddx + \frac{j^2-1}{4x^2} \pm i \frac{\eta}{x^{3/2} \sqrt{r_H}} + \OO\left((x r_H)^{-1}\right) - \omega^2) \right] \psi = 0,
\end{equation}
with
\begin{equation} \label{eta-definition}
\eta = \frac{(j^2 - 1) - 3l(l+1)}{6 \sqrt{2}}.
\end{equation}
Further setting $\rho = \omega x$,
\begin{equation} \label{wave-equation-rho}
\left[ -\dddrho + \frac{j^2-1}{4 \rho^2} \pm i \frac{\eta}{\rho^{3/2}\sqrt{\omega r_H}} + \OO\left(\frac{1}{\rho \omega r_H}\right) - 1 \right] \psi = 0
\end{equation}
Integrating the subleading terms twice in $\rho$ shows they can at
worst give corrections where $\delta \psi / \psi$ scales as a power of $\rho/\omega r_H$.
Then in the $\omega \to i \infty$ limit with
$\rho \ll 1$, one may neglect this term as well as the constant term $1$
and write simply
\begin{equation}
\left[ \dddrho + \frac{1-j^2}{4 \rho^2} \right] \psi = 0.
\end{equation}
So when $\rho \ll 1$ there are two possibilities for the asymptotics: 
\begin{equation} \label{small-rho-power-law}
\psi(\rho) \sim \rho^{(1 \pm j) /2}.
\end{equation}
It turns out that both possibilities occur and 
we can pick two linearly independent solutions $\psi_\pm$ by fixing their
rotation properties (at least for non-integer $j$):
\begin{equation} \label{solution-rotation}
\psi_\pm(e^{i \theta} \rho) = e^{(1 \pm j) i \theta / 2} \psi_\pm(\rho).
\end{equation}
On the other hand, for $1 \ll \rho$,
\eqref{wave-equation-rho} is simply
\begin{equation}
\left[ \dddrho + 1\right] \psi = 0
\end{equation}
and so the solutions are asymptotic to free waves $e^{\pm i \rho}$.  The issue is now to
interpolate between these two regions.  To leading order in $\omega$ this
can be done by neglecting the $1/\sqrt{\omega}$ term in \eqref{wave-equation-rho};
one then finds that $\psi_\pm$ are asymptotic to $\sqrt{\rho} J_{\pm j/2}(\rho)$ in the region
$\rho \ll \abs{\omega} r_H$, and since this region overlaps with the region 
$1 \ll \rho$ in the limit of large $\omega$, one can match the known 
asymptotics of Bessel functions to the free wave propagation.  
This was the approach taken in \cite{Motl:2003cd} and it gives for $1 \ll \rho$
\begin{equation} \label{solution-asymptotics}
\psi_\pm(\rho) \sim \cos(\rho - (1 \pm j)\pi / 4) + \err.
\end{equation}
From \eqref{eta-definition}, \eqref{wave-equation-rho} we see that the \err correction to
\eqref{solution-asymptotics} should be proportional to $(j^2-1) - 3l(l+1)$; in the case
$j=2$ this dependence 
was also found in \cite{MaassenvandenBrink:2003as}, where in addition the exact numerical coefficient
was evaluated.

\bibliography{physics}

\end{document}